\documentstyle[12pt,epsf]{article}
\newcommand{\be}{\begin{equation}}
\newcommand{\ee}{\end{equation}}
\newcommand{\ipb}{\int^1_0d\beta}

\newcommand{\idb}{\int_0^1d\beta\nu}
\newcommand{\iub}{\int_0^1\frac{d\beta}{\nu}}
\newcommand{\vx}{\vec{\xi_1}}
\newcommand{\vet}{\vec{\xi_2}}

\begin{document}

\begin{center}
{\Large\bf QCD string in the baryon}
\vspace*{0.5cm}

\large
Yu.S.Kalashnikova\footnote[1]{e-mail:yulia@a3.ph.man.ac.uk,
yulia@vxitep.itep.ru}\\[2mm]
{\normalsize\it Department of Physics and Astronomy,\\
University of Manchester\\
Manchester M13 9PL, UK\\[2mm]
Institute of
Theoretical and Experimental Physics\\
117259, Moscow, Russia}\\[2mm]
{\normalsize and}\\[2mm]
A.V.Nefediev\footnote[2]{e-mail:nefediev@vxitep.itep.ru}\\[2mm]
{\normalsize\it Institute of
Theoretical and Experimental Physics\\
117259, Moscow, Russia}
\end{center}

\begin{abstract}
The QCD--motivated constituent string model is extended to consider
the
baryon. The system of three quarks propagating in the confining
background
field is studied in the Wilson loop approach, and the effective
action is
obtained. The resulting Lagrangian at large interquark distances
corresponds
to the Mercedes Benz string configuration. Assuming the quarks to
be heavy
enough to allow the adiabatic separation of quark and string
junction
motion and using the hyperspherical expansion for the quark
subsystem we
write out and solve the classical equation of motion for the
junction. We
quantize the motion of the junction and demonstrate that the
account of
these modes leads to the effective \lq\lq swelling" of the baryon
in comparison
with the standard potential picture. We discuss the effects of the
finite
gluonic correlation length which do not affect the excited states
but
appear to be substantial for the baryonic ground state, reducing the
\lq\lq swelling" considerably and leaving room to the short range
Coulomb force
in the three quark system.
\end{abstract}
\smallskip

\hspace*{1cm}PACS numbers: 12.38.Aw, 12.38.Lg

{\parindent=0cm\large\bf 1\hspace{0.8cm}Introduction}
\bigskip

A reasonable model for quark confinement is expected to describe on
equal footing the simplest quark bound states with zero triality,
$i.e.$ $q\bar q$
mesons and $qqq$ baryons. While in the naive potential pair-wise
picture such
uniform description is easily achieved by introducing colour factors,
in the
more sophisticated and more ambitious QCD-motivated approaches one
should
take care of the underlying dynamics of gluonic degrees of freedom
when attempting to construct the effective model for hadronic
constituents.
At least the non-local gauge invariance should be respected if the
quark
system is placed into a nontrivial QCD background $B_{\mu}$.

In the mesonic sector the object $\bar{\psi}(y)\psi(x)$ becomes
gauge--invariant if it is supplied with the phase factor (or
parallel
transporter) $\Phi$, so that the colour singlet $q\bar q$ state is
given as
\be
\Psi(x,y) =\psi^{\alpha}(x)\Phi^{\beta}_{\alpha}(y,x)
\bar{\psi}_{\beta}(y)\;,
\label{1}
\ee
where
\be
\Phi^{\beta}_{\alpha}(y,x)=(P{\rm exp}\;ig\int_y^x B_{\mu}
dz_{\mu})^{\beta}_{\alpha}\;.
\label{2}
\ee
Here $\alpha$ and $\beta$ are the fundamental colour indices,
$B_{\mu}=B_{\mu}^a t^a$ is the
background gluonic field and $P$ stands to order the Gell-Mann
matrices
$t^a$ along the contour of integration. The only way to form the
gauge--invariant state of three quarks in the fundamental colour
representation
is to introduce the so-called string junction which serves for
coupling three
quarks into a colour singlet. So the baryonic counterpart of the
state
(\ref{1}) is presented as
$$
\Psi(x_1x_2x_3,x_0)=\hspace{8cm}
$$
\be
=\psi^{\alpha}(x_1)\psi^{\beta}(x_2)\psi^{\gamma}(x_3)
\Phi^{\alpha'}_{\alpha}
(x_0,x_1)\Phi^{\beta'}_{\beta}(x_0,x_2)\Phi^{\gamma'}_{\gamma}
(x_0,x_3)
\varepsilon_{\alpha'\beta'\gamma'}\;,
\label{3}
\ee
where $\varepsilon_{\alpha'\beta'\gamma'}$ is the antisymmetric
Levi-Civita tensor. The representation (\ref{3}) actually means,
that, in contrast to the $q\bar q$ system, the $qqq$ one
is to be treated as the simplest multiquark system containing a new
object --- a string junction with its own dynamics.

In the nonrelativistic approach confinement is reduced to the
linear potential acting between the constituents, and the usual
assumption
about the motion of the junction is that the system moves in such
a way
that the sum of distances between the junction and the quarks is
minimal
(the so-called Torricelli point) [1,2]. This assumption, though being
quite natural in the potential approach,
looks somewhat artificial if we share the common belief
that QCD at large distances is a kind of string theory and the
gluonic
degrees of freedom manifest themselves as the string ones.

The latter idea was put on to phenomenological grounds in the
flux-tube
model [2] motivated by the strong-coupling expansion of QCD.
The ordinary mesons in the flux-tube model are viewed as $q\bar q$
pairs connected
by the (nonrelativistic) string in its ground state, and the
vibrational
modes of the string  correspond to hybrid mesons. The most recent
developments in the flux-tube model are given in [3].

Another approach is based on the stochastic picture of confinement
and the Vacuum Background Correlators Method [4]. It is assumed that
the
background field configurations ensure the area law for the
Wilson loop operator
giving rise to the straight-line string configuration corresponding
to the
ordinary $q\bar q$ mesons [5], and the perturbations above the
background are
responsible for the transverse motion of the string describing
hybrid
excitations. The phenomenological implications of such a model for
hybrids
were discussed in [6].

Whatever picture of string confinement one adopts, it is clear that
in the string-type language the Mercedes Benz string configuration
(\ref{3}) for the baryon means
that the junction is a special and quite distinguishable point of the
baryonic string. Its motion might be responsible for  special types
of string excitations which are absent in the simple $q\bar q$ string
and
reveal themselves in systems with more complicated colour structure.
This idea was first suggested in [2], and the motion of the junction
as
an independent degree of freedom was studied in [7]. Here we continue
these
studies and present more detailed quantitative analysis of the
spectrum
and wave function of a baryon containing four constituents
rather than
three.
\vspace{1cm}

{\parindent=0cm\large\bf 2\hspace{0.8cm}Baryonic area law and
effective
action for the $qqq$ system}
\bigskip

Our starting point is the Vacuum Background Correlators Method.To
obtain the effective action we write the Green function of the $qqq$
system
in the Feynman--Schwinger representation [8] (in Euclidean space):
$$
G(x_1x_2x_3,x_0;y_1y_2y_3,y_0)=\int_0^{\infty}ds_1\int_0^{\infty}ds_2
\int_0^{\infty}ds_3\int Dz_1Dz_2Dz_3
$$
\be
\times{\rm exp}(-{\cal K})\cdot<W>_B\;,
\label{4}
\ee
where
\be
W=\varepsilon_{\alpha\beta\gamma}\Phi_{\alpha'}^{\alpha}(\Gamma_1)
\Phi_{\beta'}^{\beta}(\Gamma_2)\Phi_{\gamma'}^{\gamma}(\Gamma_3)
\varepsilon^{\alpha'\beta'\gamma'}\;,
\label{5}
\ee
$$
{\cal K}=\sum^3_{i=1}(m^2_is_i+\frac{1}{4}\int_0^{s_i}
\dot{z}^2_i(\tau)d
\tau)\;,
$$
with boundary conditions $z_i(0)=y_i,\;z_i(s_i)=x_i$. The paths
$\Gamma_i$
in $W$ run over the trajectories of the quarks, and brackets $<{}>_B$
mean
averaging over background field configurations.

The Wilson loop operator (\ref{5}) can be represented in the form
containing
only closed Wilson loops in the fundamental repesentation by means of
the relation following from the unimodularity condition for the group
SU(3):
\be
1=\frac{1}{3!}\varepsilon_{\alpha_1\alpha_2\alpha_3}\varepsilon^
{\beta_1
\beta_2\beta_3}\Phi^{\alpha_1}_{\beta_1}(\Gamma_0)\Phi^{\alpha_2}_
{\beta_2}
(\Gamma_0)\Phi^{\alpha_3}_{\beta_3}(\Gamma_0)\;,
\label{6}
\ee
where $\Gamma_0$ is an orbitrary open path (the same for all three
$\Phi$'s
in (\ref{6}))
connecting points $x_0$ and $y_0$. With inserted unity (\ref{6})
the Wilson loop (\ref{5}) takes the form
\be
\begin{array}{c}
W=SpW_1\ SpW_2\ SpW_3-Sp(W_1W_2)\ SpW_3-\\
{}\\
-Sp(W_3W_1)\ SpW_2-Sp(W_2W_3)\ SpW_1+\\
{}\\
+Sp(W_1W_2W_3)+Sp(W_3W_2W_1)\;,
\label{7}
\end{array}
\ee
where $(W_i)^{\alpha}_{\beta}$ is the ordered exponent along the
contour formed by the
path $\Gamma_i$ of the $i$-th quark and $\Gamma_0$ (see Fig.1).
In what follows $\Gamma_0$ will
obviously be interpreted as the path of the junction.

ince the relation (\ref{6}) is the identity for the given background,
one
may
rewrite $W$ as
\be
W=\frac{\int Dz_0 W}{\int Dz_0}\;,
\label{8}
\ee
with $z_0(\tau)$ running along the contour $\Gamma_0$,
introducing explicitly the integration over the junction
trajectories.

To average the Wilson loop configuration (\ref{7}) over the
background we
use the cluster expansion method [4] generalized in [9] for the
case of
several Wilson loops or for the loops with self-intersections.
It was demonstrated in [9] that under the assumption of existence
of
the finite gluonic correlation
length $T_g$ for the stochastic ensemble of background fields
the generalized
area law asymptotic can be obtained. For the configuration
(\ref{7}) this
generalized area law means that the average $<W>_B$ is given as
\be
<W>_B\sim\int Dz_0\;
e^{-\sigma(S_1+S_2+S_3)}\;,
\label{9}
\ee
where $\sigma$ is the string tension in the fundamental
representation
and $S_i$
is the minimal area bounded by the contours $\Gamma_i$ and $\Gamma_0$
In accordance with (\ref{8})
we integrate over the
junction trajectories ${\;z_0}$ treating in the junction as a
degree of
freedom. The area law (\ref{9}) is held for the contours with the
average
size much
larger than $T_g$ and it is violated when the contours are nearly
co-planar.
In what follows we neglect such special configurations everywhere
apart from
Section 5 where these effects will be discussed separately.

The standard approach to the junction motion corresponds to taking
only the classical trajecory $z_0(\tau)$ in the integral (\ref{9}),
or, equivalently,
assuming $\sum^3_{i=1} S_i = min$. In the nonrelativistic potential
model the latter
condition is reduced to the assumption that the sum of distances
between
the quarks and the junction is minimal. We would like to note
here that for
the isolated single Wilson loop the area law with $S=S^{min}$
arises in the
Vacuum Backround Correlators method if one does not take into
account the
perturbations above the background. However, it follows
straightforwardly
from the cluster expansion [9] that for the configurations under
discussion
each Wilson loop in (\ref{7}) is actually averaged independently
with the
result
$S_i=S_i^{min}$. By the way, this result coincides with the one
obtained
in the $\frac{1}{N_C}$
expansion (where only the first term in (\ref{7}) survives), though
in the cluster expansion
one does not apply the $N_C\to\infty$ limit. Still one needs
additional
arguments, which are not seen at present, to impose the additional
constraint
$\sum_{i=1}^3 S_i=min$. The absence of such additional reasons
is in fact
our main
motivation to consider the string junction as a baryonic degree
of freedom.

To reduce the four-dimensional dynamics to the three-dimensional
one we
follow the procedure suggested in [5] for the $q\bar q$ system.
Namely, we
use the parametrization
\be
z_{i\mu} = (\tau, \vec{r}_i),\;\; z_{0\mu} = (\tau, \vec{r}_0)
\label {10}
\ee
synchronizing the proper times of all the particles involved and
identifying this common proper time with the physical one. After
introduction the new variables
\be
\mu_i(\tau) =
\frac{T}{2s_i}
\\\dot{z}_{i0} (\tau)\;\;\; 0\leq \tau \leq T\;,
\label {11}
\ee
the Green function (\ref{4}) may be rewritten as
\be
G = \int D\vec{r}_1 D\vec{r}_2 D\vec{r}_3 D
\vec{r}_0
D{\mu_1} D{\mu_2} D{\mu_3} 
\exp(-A[\{\mu\}])\;,
\label {12}
\ee
where the effective action takes the form
\be
A =
\int^T_0 d\tau \Biggl[\sum^3_{i=1} \biggl(\frac{m_i^2}{2\mu_i}
+ \frac{\mu_i
\dot{r}_i^2}{2} +\frac{\mu_i}{2}+ \sigma \int^1_0 d\beta_i
\sqrt{\dot{w}^2_i
w'^{2}_i -(\dot{w}_i w'_i)^2}\biggr)\Biggr]
\label {13}
\ee
with the surfaces parametrized in the Nambu-Goto form, $w_{i\mu}$
being the coordinates of the world surfaces
$(\dot{w}_{i\mu}=\frac{\partial w_{i\mu}}{\partial\tau},
\;w'_{i\mu}
=\frac{\partial w_{i\mu}}{\partial\beta})$. Assuming, as usual
reasonable
approximation, the straight--line ansatz for the minimal surfaces:
\be
w_{i\mu}=z_{i\mu}(\tau)(1-\beta_i)+z_{0\mu}(\tau)\beta_i\;,
\label {14}
\ee
with $z_{i\mu}$ and $z_{0\mu}$ given by the equation (\ref{10}),
we arrive
at the Lagrangian
\be
L=\sum_{i=1}^3\Biggl(\frac{m_i^2}{2\mu_i}+
\frac{\mu_i\dot r^2_i}{2}+\frac{\mu_i}{2}+
\sigma\rho_i\int^1_0 d\beta_i\sqrt{1+l_i^2}\Biggr)\;,
\label {15}
\ee
where
$$
\vec l_i=\frac{1}{\rho_i}[\vec{\rho}_i\times ((1-\beta_i)\dot{\vec
r}_i+\beta_i \dot{\vec r}_0)]\;,~~ \vec{\rho_i}=\vec r_i-\vec r_0\;.
$$

Since no time derivatives of the fields $\mu_i$ enter the Lagrangian,
one
may integrate over $\mu_i$ in the Green function (\ref{12}) to obtain
another representation of the Lagrangian
\be
L=\sum_{i=1}^3\Biggl(m_i\sqrt{1+{\dot{r_i}}^2}+
\sigma\rho_i\int^1_0 d\beta_i\sqrt{1+l_i^2}\Biggr)\;,
\label {16}
\ee
that can be considered as the generalization (in Euclidean space) of
the $q\bar q$ relativistic flux-tube model Lagrangian [10].

In contrast to the quark velocities $\dot{\vec{r}}_i$ the string
junction
velocity $\dot{\vec{r}}_0$ enters the Lagrangian (\ref{15}) or
(\ref{16})
only via angular velocities ${\vec l}_i$. If one neglects the terms
$l_i^2$
under the square roots, then the variable $\vec{r}_0$ becomes a
non-dynamical
one, the integration over $\vec{r}_0$ reduces to taking the
extremum in
$\vec{r}_0$, and the standard requirement [1,2] placing the
junction at the
Torricelli point
\be
\sum_{i=1}^3\rho_i=min
\label {17}
\ee
is recovered after the Wick rotation to Minkowski space.
The particular
form (\ref{15}) of the baryonic Lagrangian with neglect of angular
velocities was extensively studied in [11], and accounts for
angular velocity
dependence, still within the assumption (\ref{17}), was
performed in [12].

The presence of square root terms in the representation
(\ref{15}) makes it
impossible to express explicitly the velocities in terms of
momenta. To
linearize the problem we introduce another set of auxiliary fields
following again the suggestion of [5]. The resulting form of the
Green
function is given by
\be
G=\int\prod^3_{i=1}D\nu_i\prod^3_{i=1}D\eta_i\prod^3_{i=1}D\mu_i
\prod^3_{i=1}D\vec r_i
D\vec r_0
\exp\left(-\int_0^T d\tau L[\{\mu,\nu,\eta\}]\right),
\label{18}
\ee
where
$$
L[\{\mu,\nu,\eta\}]=
\sum^3_{i=1}\Biggl(\frac{m_i^2}{2\mu_i}
+\frac{\mu_i\dot r_i^2}{2}+\frac{\mu_i}{2}+\int^1_0
d\beta_i\frac{\sigma^2(\vec{r}_i-\vec{r}_0)^2}{2\nu_i}+
$$
\be
+\frac{1}{2}\int^1_0 d\beta_i\nu_i
(\beta_i\dot{\vec{r}}_0+(1-\beta_i) \dot{\vec r}_i)^2+
\frac{1}{2}\int^1_0 d\beta_i\nu_i\eta_i^2(\vec r_i-\vec r_0)^2+
\label{19}
\ee
$$
+\int^1_0 d\beta_i\nu_i\eta_i(\vec r_i-\vec
r_0,\beta_i\dot{\vec{r}}_0+(1-\beta_i) \dot{\vec r}_i)\Biggr)\;.
$$

The initial repesentation is readily recovered by taking the
extremal values
in the fields $\nu_i$ and $\eta_i$.
\vspace{1cm}

{\parindent=0cm\large\bf 3\hspace{0.8cm}Effective Hamiltonian
of the
heavy baryon}
\bigskip

To formulate the Hamiltonian approach starting from the Lagrangian
(\ref{19}) one should define the canonically conjugated momenta as
\be
\vec{p}_i=\frac{\partial L}{\partial \dot{\vec{r}}_i}\;,\;\;\;
\vec{p}_0=\frac{\partial L}{\partial \dot{\vec{r}}_0}
\label{20}
\ee
and perform the Legendre transformation to obtain the Hamilton
function
\be
H=\sum_{i=1}^3\vec{p}_i\dot{\vec{r}}_i+\vec{p}_0\dot{\vec{r}}_0-L\;.
\label{21}
\ee

The Lagrangian (\ref{19}) contains only terms which are
linear and quadratic in velocities,
so that the relations (\ref{20}) may be easily inverted to express
velocities in terms of momenta. The integration over the auxiliary
fields
$\mu_i,\;\nu_i$ and $\eta_i$
can be done in the Hamiltonian form of the path integral
representation as
well as in the Lagrangian form (\ref{18}).

It appears, however, that due to the many--body nature of the problem
the resulting expression for the Hamilton function is very cumbersome
and
obscuring even before taking extrema in the auxiliary fields and
even at
the classical level. This leads to the delicate problem of
appropriate
operator ordering that would inevitably arise during the
quantization
procedure. In what follows we make several simplifying assumptions.

First, we consider the case of equal quark masses. Moreover, we will
restrict ourselves to the lowest hyperspherical harmonic [13] in the
quark subsystem, the appoximation proved to be successful in
few-body systems.
Anticipating the restriction by the symmetric quark
configurations, we assume that the path integral (\ref{18}) is
saturated
by the symmetric auxiliary fields configurations with
$$
\mu_1=\mu_2=\mu_3=\mu\;,\;\;\nu_1=\nu_2=\nu_3=\nu\;,\;\;
\eta_1=\eta_2=
\eta_3=\eta\;.
$$

Defining the Jacobi coordinates as
\be
\left\{
\begin{array}{l}
\vec{r}_1=\frac{1}{\sqrt{3}}\vec{R}-\frac{1}{\sqrt{6}}\vec{\xi_1}+
\frac{1}{\sqrt{2}}\vec{\xi_2}-\frac{\zeta}{\sqrt{3}}\vec{r}\\
{}\\
\vec{r}_2=\frac{1}{\sqrt{3}}\vec{R}-\frac{1}{\sqrt{6}}\vec{\xi_1}-
\frac{1}{\sqrt{2}}\vec{\xi_2}-\frac{\zeta}{\sqrt{3}}\vec{r}\\
{}\\
\vec{r}_3=\frac{1}{\sqrt{3}}\vec{R}+\sqrt{\frac{2}{3}}\vec{\xi_1}-
\frac{\zeta}{\sqrt{3}}\vec{r}\;\;\;\;\;\;\;\;\;\;\;\\
{}\\
\vec{r}_0=\frac{1}{\sqrt{3}}
\vec{R}+\frac{1-\zeta}{\sqrt{3}}\vec{r}
\;\;\;\;\;\;\;\;\;\;\;\;\;\;\;\;\;\;\;\;
\end{array}
\right.
\label{22}
\ee
and choosing $\zeta$ from the condition of mutual cancellation of
the terms
proportional to $(\dot{\vec{R}}\dot{\vec{r}})$ we obtain the
Lagrangian
(in Minkowski space):
$$
L=-\frac{3m^2}{2\mu}-\frac{3\mu}{2}-\frac{3}{2}\ipb\nu+
\frac{1}{2}\bar{\mu}(\dot{\xi_1}^2+\dot{\xi_2}^2)+
\frac{1}{2}m_s\dot{r}^2+
\frac{1}{2}M\dot{R}^2
$$
\be
-\frac{1}{2}(r^2+\rho^2)\ipb\left(\frac{\sigma^2}{\nu}-
\eta^2\nu\right)
\label{23}
\ee
$$
+\ipb\nu\eta[(1-\beta)(\vec{\xi_1}\dot{\vec{\xi_1}}+\vec{\xi_2}
\dot{\vec{\xi_2}})+
(\zeta-\beta)(\vec{r}\dot{\vec{r}})-(\vec{r}\dot{\vec{R}})]\;,
$$
where
$$
\zeta=\frac{\ipb\nu\beta}{\mu+\ipb\nu}\;,
$$
\be
\bar{\mu}=\mu+\ipb
\nu(1-\beta)^2,\;m_s=\ipb\nu\beta(\beta-\zeta)\;,\;M=\mu+\ipb\nu
\label{24}
\ee
and $\rho=\sqrt{\xi_1^2+\xi_2^2}$ is the grand hyperspherical radius.

Another simplifying assumption is to consider the system of heavy
quarks
$(m\gg\sqrt{\sigma})$, which allows one to set
$$
\dot{\vec{\xi_1}}=\dot{\vec{\xi_2}}=\dot{\vec{R}}=0\;,\;\;\zeta=0
$$
everywhere in (\ref{23}), (\ref{24}) apart from the quark kinetic
term. The
integration over $\eta$ and $\mu$ is easily performed, yielding
$\mu=m$,
the centre of mass motion is trivially separated, and the
Hamilton function
for the heavy baryon in the centre-of-mass frame takes the form
$$
H=3m+\frac{p^2+q^2}{2m}+\frac{1}{2m_s}\left(Q^2+\frac{(\vec{Q}
\vec{r})^2}{\rho^2}\right)\hspace*{4.5cm}
$$
\be
\hspace*{4.5cm}+\frac{3}{2}\ipb\nu+\ipb\frac{\sigma^2
(r^2+\rho^2)}{2\nu}\;,
\label{25}
\ee
where $\vec{p}$, $\vec{q}$ and $\vec{Q}$ are the momenta
conjugated to the
coordinates $\vec{\xi_1}$, $\vec{\xi_2}$ and $\vec{r}$
correspondingly.

The condition $m\gg\sqrt{\sigma}$ allows to treat the system in the
adiabatic approximation separating the motion of the \lq\lq fast"
string
junction and the \lq\lq slow" quark subsystems. Following the
adiabatic
procedure we rewrite the Hamilton function (\ref{25}) as
\be
H=3m+\frac{p^2+q^2}{2m}+H_j(\rho)\;,
\label{26}
\ee
where the string junction energy $H_j(\rho)$ takes the form:
\be
H_j(\rho)=\frac{1}{2m_s}\left(Q^2+\frac{(\vec{Q}\vec{r})^2}{\rho^2}
\right)+\ipb\frac{\sigma^2(r^2+\rho^2)}{2\nu}+\frac{3}{2}\ipb\nu\;.
\label{27}
\ee

Then
the string junction energy
$H_j$ as a function of $\rho$ should be
considered as the effective potential energy in the quark subsystem,
and,
similarly, if the motion of the junction is quantized, the
eigenvalues of the
Hamiltonian (\ref{27}), being functions of $\rho$, should be
considered as
the effective adiabatic potentials responsible for the quark
interaction.

The Hamilton function (\ref{27}) of the string junction looks
relatively
simple, but the quantization is not very straightforward: care
should be
taken of the operator ordering in (\ref{27}), and usual Weyl
ordering
prescription [14] is not of the great help here.

Indeed, usually, when the interaction depends only on the
coordinates,
the ordering ambiguity is resolved in the obvious way.
The interaction we deal with here is essentially string-type,
$i.e.$ nonlocal
and velocity-dependent. The approximations we have made
allow us
to localize this feature to a large extent, but not to get rid of it:
the kinetic term in (\ref{27}) causes difficulties in the canonical
quantization.
Its containing the part proportional to $(\vec{Q}\vec{r})^2$ which
explicitly
mixes canonically conjugated coordinate and momentum is not the
whole story
--- being polynomial in both it can easily be ordered by Weyl.
The real problem is that the quantization
is to be carried out only {\it after} taking the extremum
in $\nu$ and
substituting $\nu=\nu_{ext}$ into the Hamiltonian. But the
field $\nu$ enters
the effective string mass $m_s$ (\ref{24}),
and taking extremum in $\nu$ will mix
the coordinate and momentum in a severely nonlinear way,
leaving no hope
to get the final answer for the Hamilton operator which
is of any
practical use after Weyl ordering. In what follows we
develop the
Bohr--Zommerfeld quantization procedure, which allows to avoid these
difficulties, though at the price of losing accuracy.
\vspace{1cm}

{\parindent=0cm\large\bf 4\hspace{0.8cm}String junction motion}
\bigskip

The strategy adopted in what follows is to find the eigenvalues
of the
Hamiltonian (\ref{27}) with $\nu$ as an arbitrary function
of $\beta$
and $\rho$ and then minimize each eigenvalue with respect
to $\nu$, so
that $\nu=\nu_{ext}$ will depend on junction quantum numbers.

We start with the classical equation of motion for the junction which
follows from the Hamilton function (\ref{27}) in the spherically
symmetric case $L_j=0$:
\be
\frac{\ddot{z}}{1+z^2}-\frac{z\;\dot{z}^2}{(1+z^2)^2}+\omega^2z=0\;,
\label{28}
\ee
where $z=r/\rho$, dots stand for the time derivatives,
and the shorthand
notation
$$
m_s\omega^2=\sigma^2\int_0^1\frac{d\beta}{\nu}\;,\;\;i.e.\;\;\;\;
\omega=
\sigma\sqrt{\frac{\int_0^1\frac{d\beta}{\nu}}{\ipb\nu\beta^2}}
$$
is introduced. The equation of motion (\ref{28}) can be
integrated with
the result:
\be
\omega(\tau-\tau_0)=-\frac{1}{\sqrt{1+a^2}}F
\left(arccos\frac{z}{a}\;,\;
\frac{a}{\sqrt{1+a^2}}\right)\;,
\label{29}
\ee
where $F(\psi,p)$ is the elliptic integral of the first
kind (see Appendix A
for the details). The expression (\ref{29}) defines the
function $z=z(\tau)$
with $a$ and $\tau_0$ being the constants of integration.

Since the velocity $\dot{z}$ vanishes at $z=\pm a$ (as
it may be seen from
eq.$(A.2)$) the motion is finite and periodic, and the adiabatic
invariant
$$
I=\frac{1}{2\pi}\oint\vec{Q}d\vec{r}=\frac{2}{\pi}\int_{0}
^{r_{max}}Q(r)dr
$$
can be calculated with the help of the eqs.$(A.6)$, $(A.7)$:
\be
I=\frac{2}{\pi}m_s\omega\rho^2J(a)\;,\;\;J(a)=\int_0^adz\sqrt{\frac
{a^2-z^2}{1+z^2}}\;.
\label{30}
\ee

The Bohr--Zommerfeld quantization rule for the three-dimensional
radial
motion gives
\be
I=2n+\frac{3}{2}\;,\;\;\;n=0,1,2,\ldots\;\;,
\label{31}
\ee
so that the expression
\be
J(a_n)=\frac{\pi(n+\frac{3}{4})}{m_s\omega\rho^2}
\label{32}
\ee
defines $a_n$ as the function of $\rho$ for each radial quantum
number $n$
(\ref{31}). The energy of radial motion is given by
\be
V_n(\rho)=\frac{1}{2}m_s\omega^2\rho^2(a_n^2+1)+\frac{3}{2}\ipb\nu\;.
\label{33}
\ee

The expressions (\ref{32}) and (\ref{33}) solve the problem of
quasiclassical
quantization defining the eigenenergies as the functions
of $\rho$ for
each $n$. It is impossible in general to write out the analytical
expression
for the function $V_n(\rho)$. However some limiting cases
are tractable.

First we note that the quantity $a$ has the meaning of the
(dimensionless)
amplitude for the classical motion, so that the case of small
$a$ $(a\ll 1)$
corresponds to large $\rho$ $(\rho\gg r)$ and {\it vice versa}.
In the case of asymptotically large $\rho$ ($\rho\gg\frac
{1}{\sqrt{\sigma}}$) one
expects the string modes to die out. It follows from the expression
$(A.2)$
for the junction velocity that the limit $a\ll 1$ means the limit
$\dot{r}\ll 1$, $i.e.$ the \lq\lq freezing" of the junction
around the
Torricelli
point (\ref{17}).
Indeed, with the asymptotic $(A.8.a)$ one has
\be
V_n(\rho\to\infty)=\frac{1}{2}\sigma^2\rho^2\iub+\frac{3}{2}\idb
+\delta
V_n(\rho)\;,
\label{34}
\ee
where
$$
\delta V_n(\rho)=\omega \left(2n+\frac{3}{2}\right)\;.
$$

The extremum in $\nu$ can be easily taken in (\ref{34}) with the
result
\be
V_n(\rho\to\infty)=\sigma\rho\sqrt{3}+\frac{3}{\rho}\left(2n+
\frac{3}{2}\right)\;,
\label{35}
\ee
so that the spectrum (\ref{33}) becomes the spectrum of the
harmonic oscillator,
which in the leading order in $\rho$ gives the potential-type linear
confinement.

The actual coefficient $\sqrt{3}$ of the confinement term
in (\ref{35})
differs slightly from the one obtained by taking the lowest
hyperspherical
harmonic of the interaction (\ref{17}). This discrepancy
(about 15\%) stems
from setting all $\nu_i$ and $\eta_i$ equal to each other before
quantization and hyperspherical expansion, so that our results
should be
viewed as the variational estimation of the path integral (\ref{18}).

The opposite limiting case of small $\rho$ is less explicit.
One can only
consider $V_n$ as the function of $a_n$ for $a_n\gg 1$, what
corresponds
to small $\rho$. Taking into account the asymptotic $(A.8.b)$ we
have from
(\ref{32}) and (\ref{33}):
\be
V_n(a_n\to\infty)=\frac{\pi}{2}\left(n+\frac{3}{4}
\right)\omega[\nu]\frac{a_n}
{log\,a_n}
+\frac{3}{2}\ipb\nu\;,
\label{36}
\ee
or after taking the extremum in $\nu$
\be
V_n(a_n\to\infty)\;\sim\;\sqrt{\frac{a_n}{log\,a_n}}
\;\to\;\infty\;.
\label{37}
\ee

The energy of the string junction grows both for large
and small values
of $\rho$, and
the position of the minimum of the curve is defined from the
conditions
\be
\left.\frac{\partial V_n(\rho)}{\partial \rho}\right|_{
\begin{array}{l}
\rho=\rho_0(n)\\\nu=\nu_{ext}
\end{array}}=0\;,\;\;\;\left.\frac{\delta V_n[\nu]}
{\delta \nu}\right|_{\begin{array}{l}\rho=\rho_0(n)\\\nu=
\nu_{ext}\end{array}}
=0\;,
\label{38}
\ee
which result in the following expression for $\nu_{ext}$:
\be
\nu_{ext}=\sqrt{\frac{2\left(n+\frac{3}{4}
\right)((a^{(0)})^2+1)\sigma}
{3J(a^{(0)})}}\frac{1}{\sqrt{1-\beta^2}}\;,
\label{39}
\ee
where $a^{(0)}\approx 2.2$ is the root of the equation
\be
(a^2+1)\frac{\partial J(a)}{\partial a}-2aJ(a)=0\;.
\label{40}
\ee

Substituting $\nu_{ext}$ from the eq.(\ref{39}) into (\ref{32})
and (\ref{33})
we obtain the position of the minimum of the curve
$V_n(\rho)$ and the
value of energy in this point as
\be
\rho_0(n)=\sqrt{\frac{4n+3}{\sigma J(a^{(0)})}}
\label{41}
\ee
and
\be
V_n(\rho_0)=\sqrt{\frac{3\pi^2\sigma}{2}\left(n+\frac{3}{4}\right)
\frac
{(a^{(0)})^2+1}{J(a^{(0)})}}
\label{42}
\ee
correspondingly.

The results for the junction spectrum are obtained by the
quasiclassical
procedure and hence are valid for $n\gg 1$, apart from the case
of large
$\rho$ (\ref{35}) where the quasiclassical spectrum of the
harmonic oscillator
is exact. The quasiclassical results are, however, known to be rather
accurate far beyond the formal limits of application. So we dare to
apply the
results to the low-lying junction excitations, and even to the ground
state. Substituting $n=0$ into the expressions (\ref{41}) and
(\ref{42})
we find for ground state
\be
\rho_0(n=0)\approx \frac{1.03}{\sqrt\sigma}\;,\hspace*{0.5cm}
V_0(\rho_0)\approx 4.8\sqrt\sigma\;.
\label{43}
\ee

It is seen from (\ref{43}) that as the dynamics of the junction
is defined
by the string-type interaction, the typical distances $\rho_0$ are
determined by the confinement scale $\frac{1}{\sqrt\sigma}$.
Allowing for quark
motion will result only in the corrections to (\ref{43})
of order of $\frac{\sqrt\sigma}{m}$
and we arrive in such a way to a rather shocking conclusion:
the size of a heavy symmetric baryon is practically independent
of the
quark mass and is essentially defined by the string tension. This
conclusion
holds even for the ground state, and the zero oscillations of
the string
are responsible for the effective \lq\lq swelling" of the baryon.

To conclude this Section we briefly comment on the case of
$L_j\neq 0$. One
might naively expect that, as soon as the kinetic term for
orbital motion
in the Hamiltonian (\ref{27}) has \lq\lq normal"
quasi-nonrelativistic
form, then
at least for large orbital momenta $(L_j\gg 1$, $L_j\gg n_r)$
the radial
motion can be neglected, and the resulting energy $V(\rho)$
will exhibit
regular behaviour at small $\rho$. It is clear, nevertheless,
that the
case of $L_j\neq 0$ is to be considered along the same pattern, and
zero oscillations in radial motion will again result in the effective
\lq\lq swelling"
of the system. We have not carried out the quantization in this
case, mainly because for $L_j\neq 0$ there are no
{\it a priori} reasons for
restricting by the lowest hyperspherical harmonic in the quark
subsystem.
On the contrary, we expect the total wave function of the baryon
with
nonzero total orbital momentum to acquire considerable
contributions from
configurations with nonzero interquark orbital momenta.
\vspace{1cm}

{\parindent=0cm\large\bf 5\hspace{0.8cm}Mass spectrum
of heavy baryon}
\bigskip

The Schroedinger equation for the quark subsystem is obtained
from the
Hamiltonian (\ref{26}) with replacing the string junction
energy $H_j(\rho)$
by the quantized energy of the junction radial motion $V_n(\rho)$.
If only
the lowest symmetric hyperspherical harmonic is taken into
account, the
effective one--dimensional equation
\be
\left[3m-\frac{1}{2m}\frac{d^2}{d\rho^2}
+V(\rho)\right]\varphi_0(\rho)=E\varphi_0
(\rho)
\label{44}
\ee
with the standard boundary conditions
$$
\begin{array}{l}
\varphi_0(\rho=0)=0\\
\varphi_0(\rho\to\infty)\to 0
\end{array}
$$
and effective interquark potential
\be
V(\rho)=\frac{15/4}{2m\rho^2}+V_n(\rho)
\label{45}
\ee
defines the mass spectrum of the system (the necessary details
of the
hyperspherical expansion are listed in the Appendix B). Note
that each
adiabatic potential $V_n(\rho)$ gives rise to the whole family of
excitations in the quark subsystem in accordance with the equation
(\ref{44}).

The effective potential (\ref{45}) is shown at Fig.2 for
various values
of $n$ and
quark mass $m$. The junction eigenenergy $V_n(\rho)$
diverges at small $\rho$ only logarithmically, and the
centrifugal barrier
dominates the region of small $\rho$.
Nevertheless, since the quarks are heavy
$(m\gg \sqrt{\sigma})$
the minimum of the potential $V(\rho)$ coincides with good accuracy
with the minimum of the curve $V_n(\rho)$ given by the eq.(\ref{41}).

To find the eigenenergies we decompose the effective
potential (\ref{45})
around the point $\rho_0$ up to the terms $\sim
(\rho-\rho_0)^2$ using the
extremal value of $\nu$ given by the expression (\ref{39}).
The resulting
spectrum is reduced to the spectrum of the harmonic
oscillator with the
eigenenergies
\be
\varepsilon_{n_q}=V(\rho_0)+\omega_q\left(n_q+\frac{1}{2}\right)\;,
\label{46}
\ee
where
$$
\omega_q=\sqrt{\frac{V''_n(\rho_0)}{m}}\;.
$$

Substituting $a^{(0)}\approx 2.2$ from (\ref{40}) one has
$$
V(\rho_0)\approx\sqrt{\frac{3\pi^2\sigma}{2}\left(n+
\frac{3}{4}\right)
\frac{(a^{(0)})^2+1}
{J(a^{(0)})}}\approx 4.8\sqrt{\sigma\left(1+\frac{4}{3}n\right)}\;,
$$
\be
V''(\rho_0)\approx\frac{\pi\sigma^{3/2}}{4{(a^{(0)})}^2}\sqrt{
\frac{J(a^{(0)})(1+(a^{(0)})^2)^3}{2
\left(1+\frac{4}{3}n\right)}}\approx\frac{2.7\sigma^{3/2}}{\sqrt{1
+\frac{4}{3}n}}
\label{47}
\ee
with the ground state energy $(n=n_q=0)$:
\be
E_{\rm ground}=3m\left[1+1.6\frac{\sqrt{\sigma}}{m}+0.9\left(
\frac{\sqrt{\sigma}}{m}\right)^{3/2}+O\left(\left(\frac{
\sqrt\sigma}{m}
\right)^2\right)\right]\;.
\label{48}
\ee

The estimate (\ref{48}) for the ground state baryonic
energy should be
confronted with the eigenenergy derived from the standard
approach to
the junction motion, governed by the interaction (\ref{17}),
with the
junction frozen at the Torricelli point. In accordance with
the discussion
of the previous Section, the reliable estimate for this case
can be
obtained with the first term of the interaction (\ref{35}), and
instead of (\ref{48}) we arrive at
(see also [11] for more accurate estimates)
\be
\hat{E}_{\rm ground}=3m\left[1+1.5\left(\frac{\sqrt{\sigma}}
{m}\right)^{4/3}+O\left(\left(\frac{\sqrt\sigma}{m}\right)^{8/3}
\right)\right]\;.
\label{49}
\ee

Comparing the spectra (\ref{48}) and (\ref{49}) one observes
that they
differ parametrically: the excess over $3m$ in (\ref{49}) is
of order
of
$\frac{\sigma^{2/3}}{m^{1/3}}$
as it should be on the dimentional reasons for the linear
potential, while
in (\ref{48}) it is of order of $\sqrt{\sigma}$, reflecting
the string
nature of the interaction.

Moreover, for the potential case (\ref{17}), (\ref{49}) the
position of
the potential minimum depends on quark mass --- for the ground
state one has:
\be
\hat{\rho}_0=\frac{1.3}{(\sigma m)^{1/3}}\;,
\label{50}
\ee
which tends to zero with increasing quark mass, in contrast to the
effective \lq\lq swelling"
exhibited by the string approach with $\rho_0$ defined
from the expression (\ref{41}).

Before proceeding further with the analysis of the ground
state we introduce
short range effects, the first of which is to account
for perturbations
above the
confining background field. The effect of perturbative
gluons is twofold.
The transverse propagating gluonic excitations give rise
to the string
vibrations and would result in hybrid baryonic excitations
similar to the
mesonic ones [6]. The Coulombic  gluons lead to the pair-wise
interaction of the form
\be
V^{coul}(r_{ij})=\frac{1}{2}\sum_{i\ne j}^{}C\;\frac{\alpha_s}{\mid
\vec{r}_i-\vec{r}_j\mid}\;,
\label{51}
\ee
where the colour factor $C$ is equal to $-\frac{2}{3}$, so that
the interaction (\ref{51}) gives rise to the attraction
between the quarks.

Averaging the interaction (\ref{51}) over the lowest hyperspherical
harmonic with the help of the eq.$(B.4)$ we arrive at the effective
Coulombic potential
\be
V^{coul}_{00}(\rho)=-\gamma\;\frac{\alpha_s}{\rho}\;,\;\;
\gamma=\frac{16\sqrt{2}}{3\pi}\approx 2.4\;,
\label{52}
\ee
which should be added to the effective interaction (\ref{45}).

It appears, however, that for \lq\lq realistic" values of $\alpha_s$
(in actual calculations we take $\alpha_s=0.3$) and quark
masses the inclusion of the interaction (\ref{52}) leads only
to minor
numerical changes in the estimates (\ref{48}) for the energy and
(\ref{43}) for the radius of the ground state, and the
higher the state
the less it is affected.

There is another short--ranged effect, namely, effect of the finite
gluonic correlation length $T_g$. Indeed, lattice
calculations give the
estimate $T_g\approx 0.2 - 0.3 fm$ [15]. Such a value of $T_g$ is
comparable with the typical interquark distances,
and should be somehow
taken into account.

As was mentioned in Section 2, the area law (\ref{9}) is violated
for the specific configurations when all three surfaces
or at least two
of them are close to each other, $i.e.$ the average
distance between them is
of order or less than $T_g$. The configurations we should
be worried about
in our spherically symmetric baryon are the ones with
all the interquark
distances less than $T_g$.

To handle the problem we first consider the
case of coinciding quarks
trajectories. Assuming the surfaces to be flat enough we follow the
procedure suggested in [9] to calculate the average
of the complicated
Wilson loop
configurations. Namely, one should find all possible ways to
close the open contours under consideration, and rewrite the Wilson
loop configuration, using the Fierz identities, as a sum over
colour charges in all possible
irreducible representations of the SU(3) colour group which
propagate along the
coinciding contours, with the coeficients given by the
dimensions of the
representations. Similar results were also derived for the 1+1 gauge
theories [16] on the basis of completely different formalism of
Migdal--Makeenko loop equations [17].

There is no need, however, to use the whole machinery of [9],
because the
configuration (\ref{7}) is already one of such representations:
the string
junction is a colour singlet by construction, and configuration
(\ref{7})
in the case of coinciding contours corresponds to the colour singlet
propagation
along the common contour. As the string tension is obviously zero
in the singlet representation $(\sigma_0=0)$,
it means that the string simply vanishes at
zero interquark distances:
\be
<W>_B=\frac{\int Dz_0\;6e^{-\sigma_0S}}{\int Dz_0}=6\;.
\label{53}
\ee

The regimes (\ref{9}) and (\ref{53}) match smoothly to each
other at the
interquark distances $\leq T_g$, and we account for this
phenomenologically
by introducing the cutoff function
\be
f(\rho)=1-exp\left(-\frac{l(\rho)}{T_g}\right)
\label{54}
\ee
as a multiplier at $V_n(\rho)$ in (\ref{45}). Here $l(\rho)$
is some typical
interquark distance, and we choose it to be
$\frac{1}{\sqrt{3}}\rho$.
Such a drastic
cutoff of course influences the potential curve for small
excitation numbers
$n$. Indeed, now the junction energy is regular at $\rho=0$.
Moreover,
there is no minimum for $n=0$ anymore, while for large
$n$ the minimum safely survives the cutoff.
Due to numerical reasons the
minimum appears already for $n_{cr}=3$ at $T_g=0.2\;fm$.

The absence of the minimum means that there is no
\lq\lq swelling" for $n =
0,1\; {\rm and}\;2$. Such a nonexcited baryon lives in the
potential well
given to a large extent by the Coulombic and hypercentrifugal
potentials, and
its size decreases with the increase of the quark mass. For
the standard
values of quark masses and $\alpha_s$ this case is
qualitatively similar to
the case of frozen junction [11], as it is shown at Fig.3.
Still this apparent
similarity is of purely numerical origin, and the gap between
the energies remains as the consequence of zero oscillations.

To conclude we mention that as the cutoff depends only on the
correlation
length, one may find oneself in the amazing situation for large
junction excitation numbers $n\geq n_{cr}$. In the case of
large quark masses
(numerical estimate gives
$m\geq m_{cr}\approx\frac{3}{T_g}$) with the
inclusion of Coulomb force the double well potential can
appear, and the
lowest levels will be the Coulomb ones, while another
family of levels will
populate the (practically undistorted) junction curve minimum.
\vspace{1cm}

{\parindent=0cm\large\bf 6\hspace{0.8cm}Discussion and outlook}
\bigskip

To summarize, we have demonstrated that allowing
the string junction to
be an independent degree of freedom one does not arrive
at the freezing
of the junction at the Torricelli point, in contrast
to common expectations.
The eigenmodes of the string junction motion affect
the spectrum drastically
leading to the effective \lq\lq swelling" of the
heavy baryon in comparison
to the standard potential model results. The typical
size of the heavy baryon
does not depend on quark masses and is defined by the
confinement size
$\frac{1}{\sqrt\sigma}$. The standard picture is recovered only if
the effects of the
finite gluonic correlation length are taken into account: in this
case
the string does not develop itself at full scale for the lowest
baryonic
excitations, and the dynamics is actually governed by
the Coulomb force
in the three--quark system.

The origin of the effect stems from the  special type of
baryonic string excitations, and might be traced to the underlying
dynamics of nonperturbative QCD. The gluonic degrees of
freedom reveal
themselves as the junction ones, and might be responsible
for the spin
content of a baryon.

Unfortunately, (or may be fortunately for the model) the heavy
$\Omega$-type baryons are not observed yet. The excitations
we have described
are present also in the light quark sector within the string--type
confinement model, and the effects of \lq\lq swelling"
of the baryon and pushing
up the energy will survive, though maybe in less pronounced
form. For light
quarks the adiabatic approximation is not valid anymore,
and a more adequate
approach is to consider the four-body system of light
constituents with the
string junction on a par with quarks. Both constituent
masses of light quarks
and the excitation energy are of order of $\sqrt\sigma$,
and one should not
expect the quantitative features of the four-body system
to be dramatically
different from the ones of the three-body system.
As soon as the present quark models define the energy of the
state up
to the additive constant, the increase of energy due to
junction motion
is quite tolerable and might be accounted for by refitting
the constant. In a
similar way, the increase in size which is expected
to be rather modest
for light quarks might be absorbed by re-adjusting the quark model
parameters. The main effect of inclusion of the junction
motion is to change
the whole pattern of the spectrum due to the presence
of the extra constituent
which carries energy and momentum.

The most elaborated quark models [18] describe the light
baryon spectrum
rather well, although some problems remain unsolved,
notably the long-standing
problem of the Roper resonance $N(1440)$. The mass of
this state is slightly
lower than the expected mass of the first radial
excitation, its
photoproduction amplitude [19] does not fit the $qqq$
assignement, and
hybrid interpretation [20] was invoked to describe this state.
It is a
challenge for the string--junction model to describe the Roper
resonance,
together with other {\it enfants terribles} of the
baryonic family
like $\Delta(1600)$, as first string--junction excitations.
\vspace{1cm}

{\parindent=0cm\large\bf \hspace{0.8cm}Acknowledgements}
\bigskip

We would like to thank K.G.Boreskov, A.Donnachie and
Yu.A.Simonov for
enlightening discussions. One of us (Yu.S.K.) acknowledges
the warm
hospitality of the Department of Physics and Astronomy,
University
of Manchester, where this work was completed.
This research was supported
in part by the Russian Fundamental Research Foundation
(grant $N^{\underline{0}}$ 96--02--19184)
and by INTAS--93--0079.
\vspace{1cm}

{\parindent=0cm\large\bf \hspace{0.8cm}Appendix A}
\bigskip

The eq.(\ref{28}) can be rewritten in terms of
$\varphi=\dot{z}^2$,
$\varphi'=\frac{\partial \varphi}{\partial z}$ and $z$ as
$$
\frac{\frac{1}{2}\varphi'}{1+z^2}-\frac{z\varphi}{(1+z^2)^2}+
\omega^2z=0
\eqno (A.1)
$$
and integrated, giving
$$
\varphi(z)=\omega^2(1+z^2)(a^2-z^2)\;,
\eqno (A.2)
$$
where $a$ is the constant of integration.
It is more convenient to consider $z$ varying along the
whole real axis
from $-\infty$ to $+\infty$, rather then from $0$ to
$+\infty$  as it follows
from the definition of $z$, but then while quantizing we
are to take only
the odd levels (see eq.(\ref{31})).
Integrating the eq.$(A.2)$ we arrive at the eq.(\ref{29}) with
$$
F(\psi,p)=\int_0^{\psi}\frac{dy}{\sqrt{\mathstrut {1-p^2sin^2y}}}=
\int_0^{sin\psi}\frac{dy}{\sqrt{\mathstrut{(1-y^2)(1-p^2y^2)}}}\;.
\eqno (A.3)
$$

The period of motion is then calculated using the relation
$$
F(\psi+2\pi n,p)=F(\psi,p)+2n{\bf K}(p)\;,
\eqno (A.4)
$$
where {\bf K}$(p)=F(\pi/2,p)$ is the complete elliptic integral
of the first
kind, and the result is
$$
T=\frac{4}{\omega\sqrt{1+a^2}}{\bf K} \left(\frac{a}{\sqrt{1+
a^2}}\right)\;.
\eqno (A.5)
$$

To calculate the adiabatic invariant (30) we define the
momentum $Q(r)$
from the Hamilton function (27) as
$$
Q(r)=\frac{\rho^2}{r^2+\rho^2}m_s\dot{r}
\eqno (A.6)
$$
and, taking into account the relation $(A.2)$, obtain the
adiabatic invariant
(30). The function $J(a)$ entering the eq.(30) can be
expressed in terms of
the complete elliptic integral of the third kind {\bf D}$(p)$ as
$$
J(a)=\frac{a^2}{\sqrt{1+a^2}}{\bf D} \left(\frac{a}{\sqrt{1+
a^2}}\right)\;.
\eqno (A.7)
$$

The asymptotic behaviour of the function $(A.7)$ is given by
$$
J(a)=\frac{\pi}{4}a^2\;\;{\rm for}\;\;a\ll 1
\eqno (A.8.a)
$$
and
$$
J(a)=a\,log\,a\;\;{\rm for}\;\;a\gg 1\;.
\eqno (A.8.b)
$$
\vspace{1cm}

{\parindent=0cm\large\bf \hspace{0.8cm}Appendix B}
\bigskip

The hyperspherical expansion [13] implies the decomposition
of the three--body
wave function $\Psi(\vx,\vet)$ into a set of spherical harmonics
$$
\Psi(\vx,\vet)=\sum_{K=0}^{\infty}\chi_K(\rho)
u_K(\vx/\rho,\vet/\rho)\;,
\eqno (B.1)
$$
where $\rho^2=\xi_1^2+\xi_2^2$, and
$$
(\triangle_{\vx}+\triangle_{\vet})\rho^K u_K(\vx/\rho,\vet/\rho)=
(\triangle_{\vx}+\triangle_{\vet})P_K\equiv\triangle_6 P_K=0\;.
$$

With the decomposition $(B.1)$ the three--body Schroedinger
equation is
reduced to the infinite set of hyperradial equations
for the functions
$\varphi_K(\rho)=\rho^{5/2} \chi_K(\rho)$:
$$
\frac{d^2\varphi_K}{d\rho^2}+
\sum_{K'=0}^{\infty}\left(\left(2m\varepsilon-
\frac{\left(K+\frac{3}{2}\right)\left(K+\frac{5}{2}\right)}
{\rho^2}\right)
\delta_{KK'}
-2mV_{KK'}(\rho)\right)\varphi_{K'}=0\;,
\eqno(B.2)
$$
with
$$
V_{KK'}(\rho)=\int u^*_K V(\vec{r}_1,\vec{r}_2,\vec{r}_3)
u_{K'} d\Omega_6\;,
\eqno(B.3)
$$
where $V(\vec{r}_1,\vec{r}_2,\vec{r}_3)$ is
the potential energy in the
three--body system.

Truncating the set $(B.2)$ and leaving only the first
equation with $K=K'=0$
one arrives at the equation (\ref{44}).

The matix element $V^{coul}_{00}(\rho)$ of the pair--wise
potential (\ref{51})
$$
V^{coul}
(\vec{r}_{12},
\vec{r}_{23},
\vec{r}_{31})
=\frac{1}{2}\sum_{i\ne j}^{}V(\vec{r}_{ij})$$
is calculated from
$(B.3)$ as
$$
V^{coul}_{00}(\rho)=\frac{\int_0^{\pi /2}d\theta
sin^2\theta cos^2\theta
\int d\Omega_{\vec{\xi}}\int
d\Omega_{\vec{\eta}}V^{coul}(
\vec{r}_{12},
\vec{r}_{23},
\vec{r}_{31}
)}
{\int_0^{\pi /2}d\theta sin^2\theta cos^2\theta\int d\Omega_
{\vec{\xi}}\int d\Omega_{\vec{\eta}}}=-\gamma\frac{\alpha_s}{\rho}
\eqno(B.4)
$$
with $\gamma\approx\frac{16\sqrt{2}}{3\pi}$.

\newpage

\newlength{\zzz}
\zzz=7.5cm
\vspace*{5cm}
\begin{figure}[h]

\newcommand{\bp}{\begin{picture}}
\newcommand{\ep}{\end{picture}}
\unitlength=1mm
\bp(150,40)
\put(15,13){\bp(71,40)%
   \put(15,10){\line(0,1){20}}
   \put(15,10){\line(3,-4){9}}
   \put(15,10){\line(-1,-1){15}}
   \put(85,10){\line(0,1){20}}
   \put(85,10){\line(3,-4){9}}
   \put(85,10){\line(-1,-1){15}}
   \multiput(15,10)(6,0){12}{\line(1,0){3}}
   \put(82,10){\line(1,0){3}}
   \put(15,30){\line(1,0){70}}
   \put(0,-5){\line(1,0){70}}
   \put(24,-2){\line(1,0){70}}
   \put(13,33){$y_1$}
   \put(17,-2){$y_2$}
   \put(-3,-8){$y_3$}
   \put(8,10){$y_0$}
   \put(86,32){$x_1$}
   \put(95,-2){$x_2$}
   \put(71,-8){$x_3$}
   \put(87,10){$x_0$}
   \put(45,13){$\Gamma_0$}
   \put(45,33){$\Gamma_1$}
   \put(54,1){$\Gamma_2$}
   \put(30,-9){$\Gamma_3$}
   \put(30,-6){$\to$}
   \put(54,-3){$\to$}
   \put(45,29){$\to$}
\ep
}
\ep
\caption{Wilson loop configuration for the baryon.}
\end{figure}
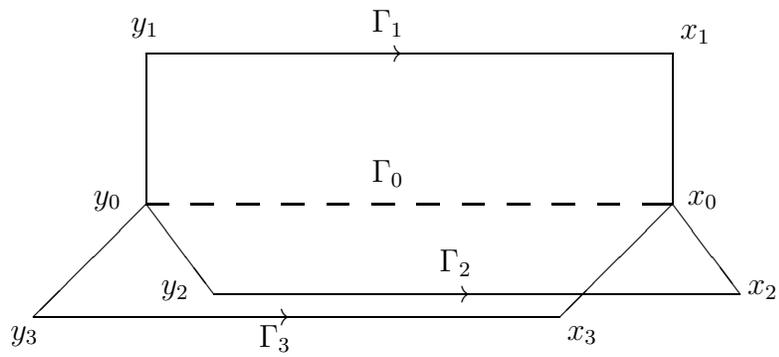

\newpage

\begin{figure}[h]
\epsfysize=\zzz
\epsfbox{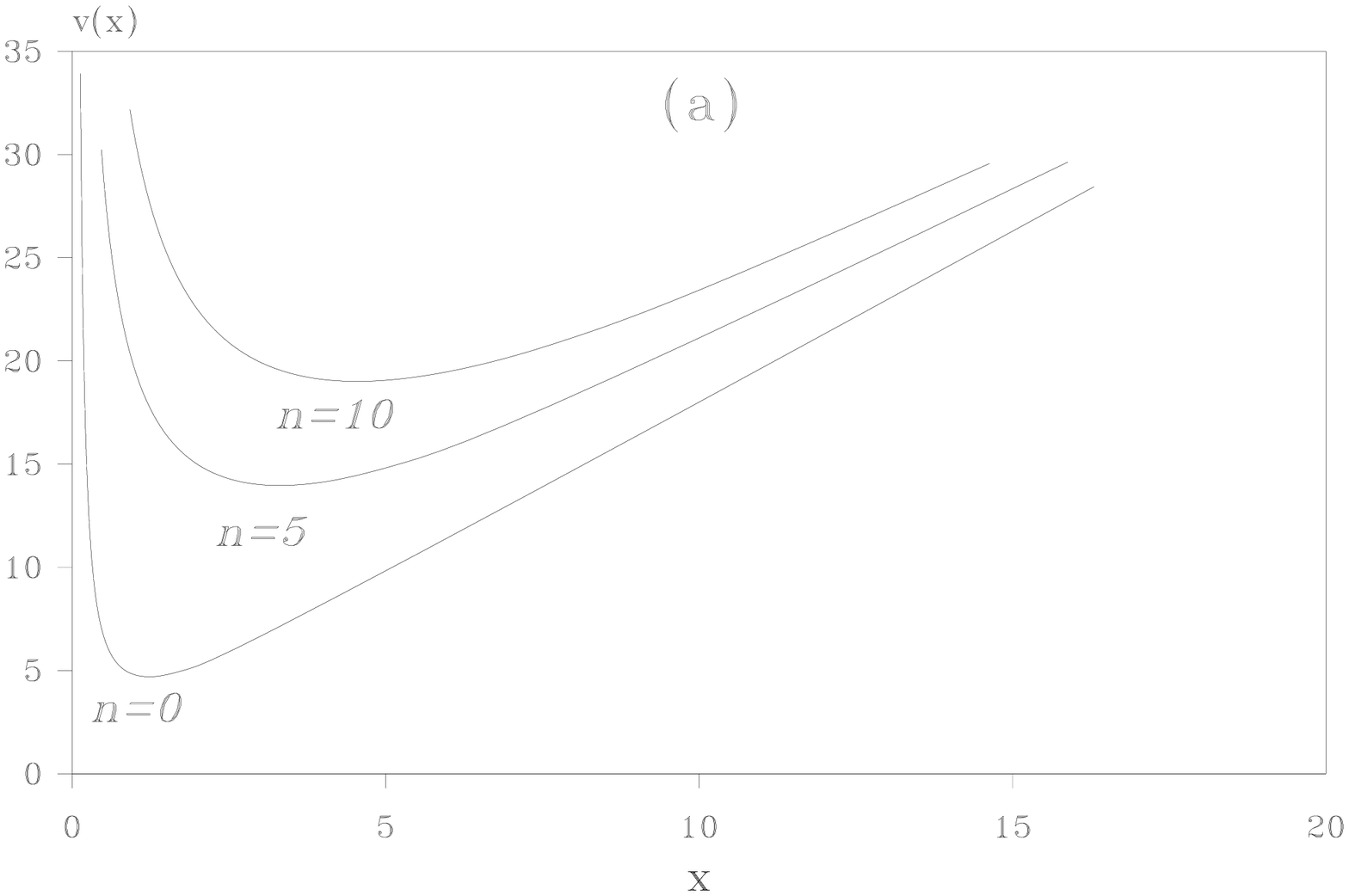}
\end{figure}
\setcounter{figure}{1}

\begin{figure}[h]
\epsfysize=\zzz
\epsfbox{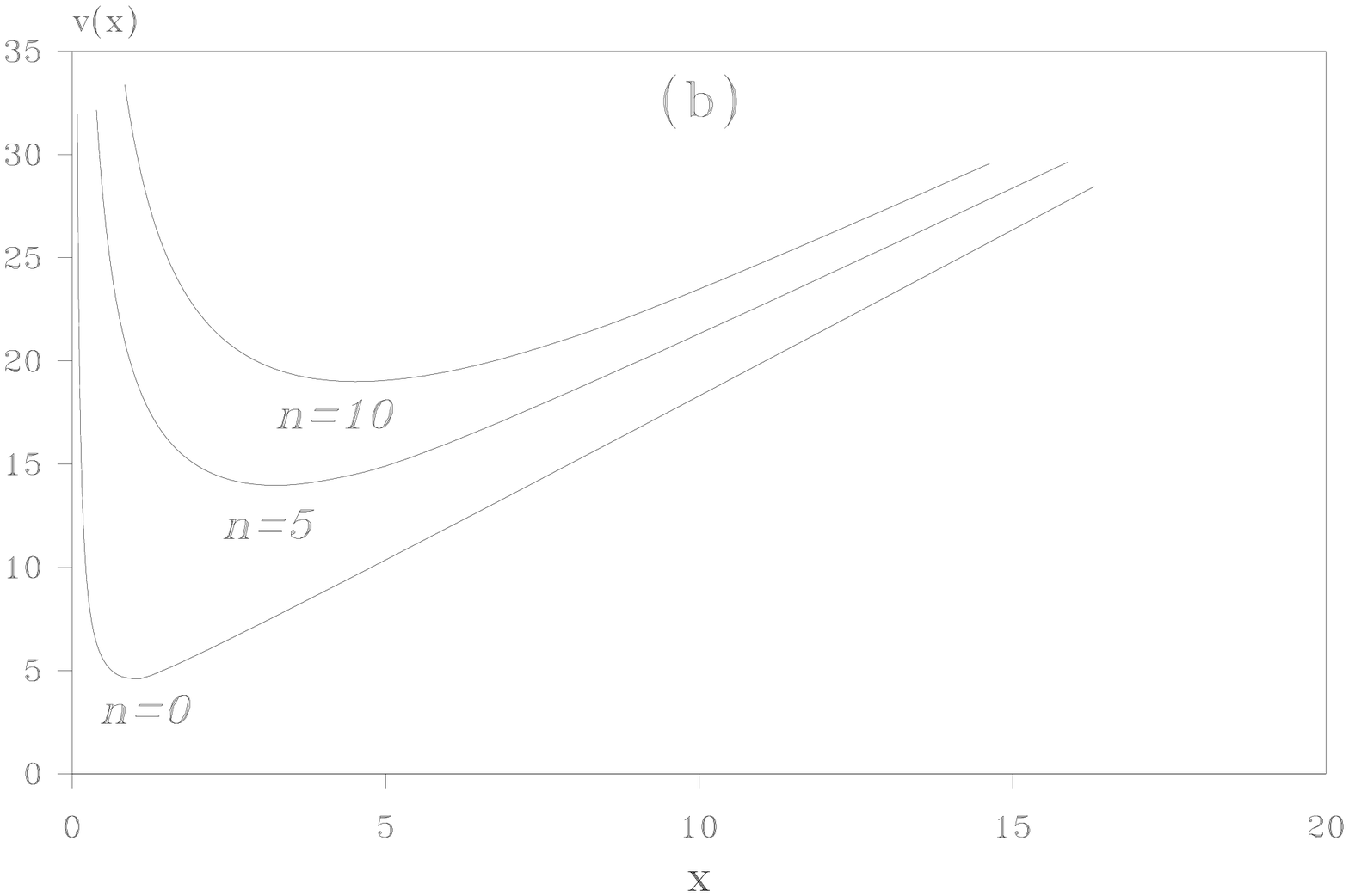}
\caption{Effective potentials v$=V_n$(x)$/\sigma^{1/2}$ as the
functions of x$=\rho\sigma^{1/2}$
with the inclusion of hypercentrifugal
barrier for $m= 1.5 GeV$ (a) and $m= 5.0 GeV$ (b).}
\end{figure}
\newpage

\begin{figure}[h]
\epsfysize=\zzz
\epsfbox{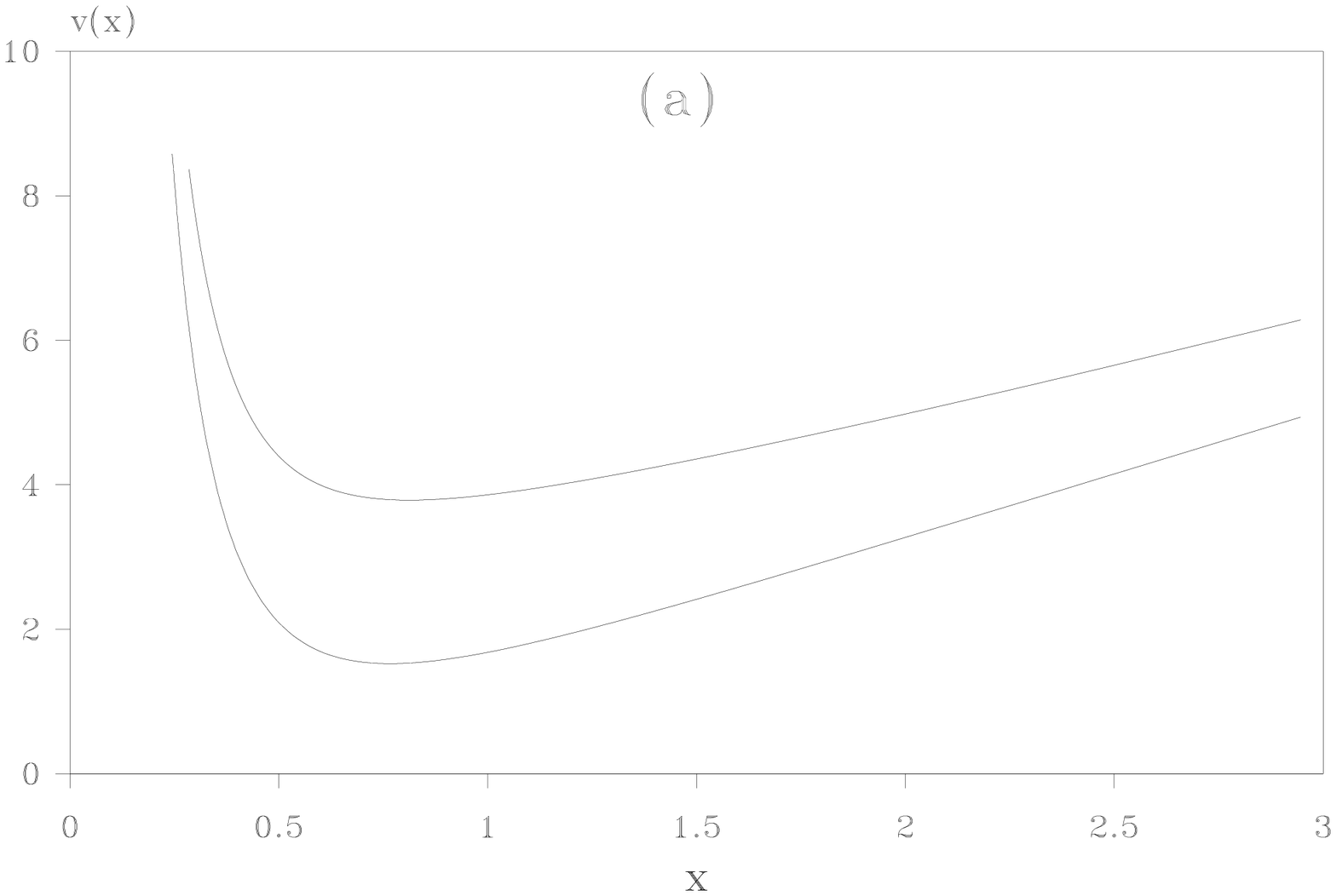}
\end{figure}
\setcounter{figure}{2}

\begin{figure}[h]
\epsfysize=\zzz
\epsfbox{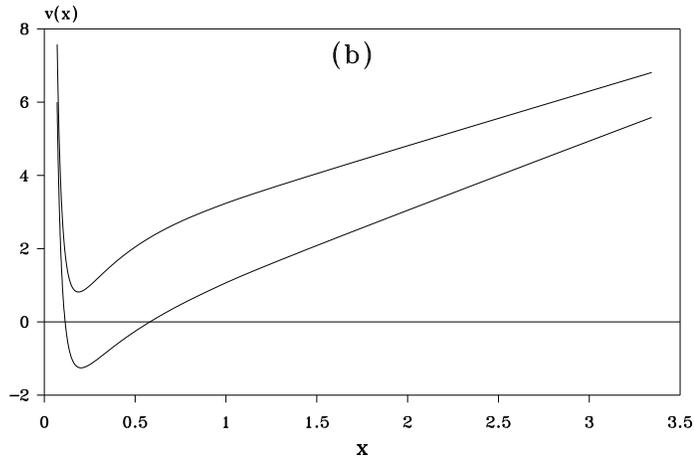}
\caption{Effective potentials v(x)=$V$(x)$/\sigma^{1/2}$,
x$=\rho\sigma^{1/2}$,
with the inclusion of hypercentrifugal
barrier and Coulomb force for $m= 1.5 GeV$ (a) and $m= 5.0 GeV$
(b); upper curve: for the string junction motion with $n=0$ and
$T_g= 0.2 fm$; lower curve: for the frozen junction.}
\end{figure}
\end{document}